\documentclass[12pt]{article}
\usepackage[left=2cm, right=2cm, top=3cm, bottom=3cm]{geometry}
\usepackage{array}
\usepackage[numbers]{natbib}
\usepackage{amsmath}
\usepackage{bm}
\usepackage{graphicx}
\usepackage{epstopdf}
\usepackage[figuresright]{rotating}
\usepackage{booktabs}
\usepackage[colorlinks=true, urlcolor=citecolor, linkcolor=citecolor, citecolor=citecolor]{hyperref}
\usepackage{bm}
\usepackage{multirow}
\usepackage{authblk}
\usepackage{fancyhdr}
\usepackage{natbib}

\title{Modeling sign concordance of quantile regression residuals with multiple outcomes}

\date{}

\author[1]{Silvia Columbu}

\affil[1]{University of Cagliari}

\author[2]{Paolo Frumento}
\affil[2]{University of Pisa}
\author[3]{Matteo Bottai}
\affil[3]{Karolinska Institute}

\begin{document}

\maketitle
\begin{abstract}
{Quantile regression permits describing how quantiles of a scalar response variable depend on a set of predictors. Because a unique definition of multivariate quantiles is lacking, extending quantile regression to multivariate responses is somewhat complicated. In this paper, we describe a simple approach based on a two-step procedure: in the first step, quantile regression is applied to each response separately; in the second step, the joint distribution of the signs of the residuals is modeled through multinomial regression. The described approach does not require a multidimensional definition of quantiles, and can be used to capture important features of a multivariate response and assess the effects of covariates on the correlation structure. We apply the proposed method to analyze two different datasets.}
\end{abstract}

\section{Introduction}
\label{sec1}
\vspace{0.5cm}
Quantile regression (e.g., \cite{KoenkBook}) can be used to assess the effect of covariates on the conditional quantiles of a univariate response variable. Extending quantile regression to multivariate outcomes, however, is not straightforward, as no natural definition of multivariate quantile is available.

Multiple outcomes of interest are common in the medical literature. For instance, clinical trials often have multiple primary endpoints, and limiting the analysis to one single endpoint may be considered undesirable. This problem is particularly important when the trialed treatment is suspected to have potentially different effects on different endpoints. For example, an anticoagulant treatment after a primary stroke may reduce the risk of a second stroke and at the same time increase the risk of additional bleeding \citep{Paciaroni}.

The analysis of multivariate quantiles has been discussed in the existing literature. An excellent review is provided in \cite{SerflingRev}. Most of the proposed methods use geometrical definitions of multidimensional quantiles that are mainly based on vector-valued ranks, and use the orientation information to identify directional quantiles in a multidimensional data cloud (\citealp{DudKolt, Chauduri, StruyfRouss, Chakraborty, HallinETal, KongMizera, LiuZuo, Chavas, Cai10, Geraci20}). Other scholars proposed joint modeling of quantiles in a likelihood framework, avoiding a mathematical definition of multivariate quantile (\citealp{Kulkarni19, PetrellaRaponi19, Alfo20}). Bayesian methods have also been described in the literature (\citealp{DrovandiPettitt11, WaldmannKneib15, Guggisberg19}). 
 
In this work, we propose a method that does not require a definition of multivariate quantile, and is implemented in two steps: at first, quantile regression is applied to each response separately; then, a suitable regression model is used to investigate the joint distribution of the signs of the residuals and their conditional association. The method aims to explore the association between multiple outcomes by investigating the joint behavior of their conditional quantiles. This approach bears some similarities with that of \cite{LiChengFine}, in which copulas are used to model the bivariate quantile-specific conditional distribution.

Our proposal requires formulating a multivariate binary response model in which the binary outcomes are defined by the sign of quantile regression residuals. In the literature, the problem of analyzing correlated binary outcomes has been tackled in different ways, which may be broadly grouped into three categories. The first category comprises methods based on generalized estimating equations that do not require specifying the joint distribution of the multivariate response (\citealp{LiangZeger, Lipsitzetal, Prentice, LuYang}). The second category consists of generalized linear mixed models (\citealp{Stiratellietal, BreslowCalyton, Dasetal, MolenberghsVerbeke}). The third and most recent category includes methods that use copulas to model the multivariate association (\citealp{MeesterMacKay, GauvreauPagano, NikoloulopoulosKarlis, Genestetal}). 

In this paper, we suggest modeling the full joint distribution of the binary responses in a simple way, using a multinomal logistic regression model. This approach does not require strong simplifying assumptions, but may become unfeasible if the response vector is high-dimensional. In many real-data settings, however, the number of responses is two or three.

The advantages in the application of the proposed method are illustrated through two different data applications. The first one considers a dataset on lung function capacity with correlated spirometrics outcome measures. The second application refers to the the National Merit Twins Study, and allows for a comparison with the quantile association method proposed by \cite{LiChengFine}.

The paper is structured as follows. In Section~2 we introduce the method. In Subsection~2.1 we present the first step of the procedure by considering univariate quantile regression models and discuss how to measure the correlation between the signs of their regression residuals. In Subsection~2.2 we define the second step by showing how the correlation between signs of quantile residuals can be allowed to depend on covariates through a multinomial regression model. In Section~3 and ~4 we present the applications.

\section{Proposed model}
\subsection{Sign-concordance of quantile regression residuals}
\label{sec:2}
We denote by $\bf{x}$ a $q$-dimensional vector of observed covariates, and by $(Y^{(1)},Y^{(2)})$ a pair of continuous response variables. Following standard quantile regression notation (\citealp{KoenkBass}), we assume the following univariate, quantile-specific linear model to hold for each response:
\begin{equation} \label{q}
Y^{(j)}_i = {\bf x}_i^T {\bm \beta}_{\tau}^{(j)} + {\bm \varepsilon}^{(j)}_i \quad j=\{1,2\}, \quad i=\{1,\ldots,n\}
\end{equation}
with $P({\bm \varepsilon}^{(j)}_i \le 0 \mid {\bf x}_i) = \tau$. In this model, $Q_{y^{(j)}}(\tau \mid {\bf x}_i^T)= {\bf x}_i^T {\bm \beta}_{\tau}^{(j)}$ represents the $\tau$-th conditional quantile of the $j$-th response, and ${\bm \beta}_{\tau}^{(j)}$ is a vector of model coefficients, $\tau \in (0,1)$. 

An estimate of the unknown quantile regression coefficients, ${\bm \beta}_{\tau}^{(1)}$ and ${\bm \beta}_{\tau}^{(2)}$, is obtained by minimizing
\begin{equation}
 \sum_{i=1}^{n} \rho_{\tau}(y^{(j)}_i - {\bf x}_j^T {\bm \beta}_{\tau}^{(j)}), \quad j=\{1,2\},
\end{equation}
where  $ y^{(j)}_i$ is a realization from $Y^{(j)}_i$, and $\rho_{\tau}(u)=(I(u \leq 0)-\tau)u$ is a loss function. We denote by 
$\hat {\bm \beta}_{\tau}^{(j)}$ the estimated regression coefficients, and by 
$\hat{\varepsilon}^{(j)}_i = y^{(j)}_i - {\bf x}^T_i \hat{{\bm \beta}}^{(j)}_{\tau}$ the corresponding quantile-specific  regression residuals. 

We define two binary indicators, namely $\omega_i^{(1)} = I(Y_i^{(1)} \leq {\bf x}^T_i {\bm \beta}^{(1)}_{\tau})$ and $\omega_i^{(2)} = I(Y_i^{(2)} \leq {\bf x}^T_i {\bm \beta}^{(2)}_{\tau})$, such that 0 and 1 indicate positive and negative residuals, respectively. We introduce the following random variable:
\begin{equation} \label{zed}
Z_i = \left\{
  \begin{array}{l l}
    1 & \quad \text{if $\omega_i^{(1)} = 0$ and $\omega_i^{(2)} = 0,$}\\
    2 & \quad \text{if $\omega_i^{(1)} = 1$ and $\omega_i^{(2)} = 1,$}\\
    3 & \quad \text{if $\omega_i^{(1)} = 0$ and $\omega_i^{(2)} = 1,$} \\
    4 & \quad \text{if $\omega_i^{(1)} = 1$ and $\omega_i^{(2)} = 0.$}
  \end{array} \right.
 \end{equation}
For a more intuitive notation, in the rest of the manuscript we will associate the labels $\{$``00'', ``11'', ``01'', ``10''$\}$ to the values $\{1,2,3,4\}$ that form the support of $Z$.  
The variable $Z_i$ carries information on the concordance of the signs of the residuals from the two regression equations defined in (\ref{q}). Discordance between signs indicates negative dependence between $Y^{(1)}$ and $Y^{(2)}$, given ${\bf x}$. Similarly, concordance between signs suggests a positive correlation.

Given the estimates $\hat {\bm \beta}_{\tau}^{(j)}$and $\hat{\varepsilon}^{(j)}_i $ we denote by
$\hat\omega_i^{(j)} = I(y_i^{(j)} \leq {\bf x}^T_i \hat {\bm \beta}^{(j)}_{\tau}) = I(\hat{\varepsilon}^{(j)}_i \leq 0)$ the estimated binary indicators of 
negative residuals. 

The observed values of $z_i$ are defined by the four possible combinations of 
$\hat\omega_i^{(1)}$, which reflects the sign of $\hat{\varepsilon}^{(1)}_i$, and $\hat\omega_i^{(2)}$, which reflects the sign of $\hat{\varepsilon}^{(2)}_i$. The sign concordance can be summarized by a measure of correlation between binary variables, i.e., the sample counterpart of the following population parameter:
\begin{equation}\label{phi1}
\phi = \text{cor}(\omega^{(1)},\omega^{(2)}) = \frac{E[\omega^{(1)}\omega^{(2)}] - E[\omega^{(1)}]E[\omega^{(2)}]}
  {\sqrt{E\left[\left(\omega^{(1)} - E[\omega^{(1)}]\right)^2\right] E\left[\left(\omega^{(2)} - E[\omega^{(2)}]\right)^2\right]}}.
\end{equation}

The equation in formula \eqref{phi1} can be rewritten as
\[\phi = \frac{F_{Y^{(1)}Y^{(2)}}(Q_{y^{(1)}}(\tau \mid {\bf x}), Q_{y^{(2)}}(\tau \mid {\bf x}))-\tau^2}{\tau(1-\tau)},\]  
where $F_{Y^{(1)}Y^{(2)}}(Q_{y^{(1)}}(\tau \mid {\bf x}), Q_{y^{(2)}}(\tau \mid {\bf x}))$ is the joint, unconditional distribution function of the responses, evaluated at the conditional quantiles.
Note that, by definition, $E[\omega^{(j)}] = P({\bm \varepsilon}^{(j)} \le 0 \mid {\bf x}) = \tau$. This holds approximately true when the sample counterparts of $\omega^{(j)}$ are used (\citealp{KoenkBass}, Theorem 3.4). The joint distribution of the quantile regression residuals signs is illustrated in Table \ref{cont_table}, where we used the notation $p_z = P(Z = z)$, $z \in \{$``00'', ``11'', ``01'', ``10''$\}$. The limits of the $\phi$-coefficient can be obtained by calculating its value in the three limiting situations summarized in Table \ref{cont_lim}.

\begin{table}[h]
\begin{center}
\small{
 \begin{tabular}{c  c | c | c | c}
& & \multicolumn{2}{ c| }{$sign \left(Y^{(2)}-{\bf x}^T {\bm \beta}^{(2)}_{\tau}\right)$} &\\
\cline{3-4}
&	&  +		    &  -	         &          \\
\hline
\multicolumn{1}{ c| }{\multirow{2}{*}{$sign \left(Y^{(1)}-{\bf x}^T {\bm \beta}^{(1)}_{\tau}\right)$}} &+   & $p_{00}$ 	& $p_{01}$	 & $p_{0*} = 1-\tau$	 \\
\cline{2-5}
&\multicolumn{1}{ |c| }{-}	& $p_{10}$	&   $p_{11}$    &	$p_{1*} = \tau$	 \\
\hline
&    & $p_{*0} = 1-\tau$		& $p_{*1} = \tau$         &	
\end{tabular}}
\end{center}
\caption{Contingency table showing the joint distribution of the signs of quantile regressions residuals. By definition, the margins are given by $P(Y^{(j)} > {\bf x}^T {\bm \beta}^{(j)}_{\tau}) = 1 - \tau$ and $P(Y^{(j)} \le {\bf x}^T {\bm \beta}^{(j)}_{\tau}) = \tau$, $j = \{1,2\}$.}
\label{cont_table}
\end{table}

Following Table \ref{cont_table}, the value of the $\phi$ statistic in \eqref{phi1} can be also expressed as:
\begin{equation}\label{phi}
\phi=\frac{p_{11}p_{00}-p_{01}p_{10}}{\sqrt{p_{0*}\times p_{1*} \times p_{*0} \times p_{*1}} }
\end{equation}
with $p_{0*}=p_{*0}=1-\tau$ and $p_{1*}=p_{*1}= \tau$.
By applying \eqref{phi} to Table \ref{cont_lim} we can derive the following limits for the correlation coefficient:
\begin{itemize}
\item  $\phi_{\text{Indep}} = 0$, independence;
\item  $\phi_{\text{Max}} = 1$, largest possible positive dependence;
\item  $\phi_{\text{Min}} =\left\{
  \begin{array}{l l}
-\tau/(1-\tau) & \quad \tau \leq 0.50 \\
 -(1-\tau)/\tau &  \quad \tau \geq 0.50
\end{array} \right\}$,
 largest possible negative dependence.
\end{itemize}

The above theoretical bounds for the $\phi$ statistic depend only on the quantile being estimated, and not on the data. The largest possible value of $\phi$ is the same as that of the Pearson's correlation coefficient. Instead, the lower limit of the coefficient is greater than $-1$, unless $\tau = 0.5$. This is a consequence of the fact that, due to the asymmetric structure of quantiles, the cells on the principal diagonal of Table \ref{cont_table} can never be simultaneously empty.

\begin{table}[h]
\begin{center}
\small{
\begin{tabular}{c  c  c | c | c | c}
& & & \multicolumn{2}{ c| }{$sign \left(Y^{(2)}-{\bf x}^T {\bm \beta}^{(2)}_{\tau}\right)$} &\\
\cline{4-5}
&	& & +	    &  -	 &          \\
\hline
\multicolumn{1}{ c| }{\multirow{8}{*}{$sign\left(Y^{(1)}-{\bf x}^T {\bm \beta}^{(1)}_{\tau}\right)$}} & \multirow{4}{*}{+}   & \multicolumn{1}{ |c| }{ Independence}& $(1-\tau)^2$  & $\tau-\tau^2$	 & \multirow{4}{*}{ $1-\tau$}	 \\
& \multicolumn{1}{ |c| }{}	& Max &$1-\tau$	    & $0$\\
& \multicolumn{1}{ |c| }{}	& Min ($\tau \leq 0.5$) &$1-2\tau$	    & $\tau$\\
& \multicolumn{1}{ |c| }{}	& Min ($\tau \geq 0.5$) &$0$	    & $1- \tau$\\
\cline{2-6}
& \multicolumn{1}{|c|}{ \multirow{4}{*}{-}}&\multicolumn{1}{ c| }{ Independence}	& $\tau-\tau^2$	    & $\tau^2$ &\multirow{4}{*}{ $\tau$}	 \\
& \multicolumn{1}{ |c| }{}	& Max &$1-\tau$	    & $0$\\
& \multicolumn{1}{ |c| }{}	& Min ($\tau \leq 0.5$) &$\tau$	    & $0$\\
& \multicolumn{1}{ |c| }{}	& Min ($\tau \geq 0.5$) &$1-\tau$	    & $2\tau-1$\\
\hline
 &  & & $1-\tau$	& $\tau$ &	 \\
\end{tabular}}
\end{center}
\caption{Joint distribution (relative frequencies) of the signs of quantile regression residuals in case of independence, perfect positive dependence (Max)
 and perfect negative dependence (Min) between the two outcomes.}
\label{cont_lim}
\end{table}

\subsection{Modeling the conditional correlation}\label{modeling}
\vspace{0.5cm}

The correlation coefficient $\phi$, that describes the association between the signs of the residuals of univariate quantile regression, is usually a function of the predictors. In particular, while the \textit{margins} of Table \ref{cont_table} are always equal to $\tau$ and $1 - \tau$, the \textit{joint} distribution of $\omega_i^{(1)} = I(Y_i^{(1)} \leq {\bf x}^T_i {\bm \beta}^{(1)}_{\tau})$ and $\omega_i^{(2)} = I(Y_i^{(2)} \leq {\bf x}^T_i {\bm \beta}^{(2)}_{\tau})$ can depend on covariates. For example, the correlation between outcomes could be larger in smokers than in non-smokers. Note that the factors influencing this association may not coincide with that used in the univariate quantile regressions.

In this paper we suggest using a multinomial logistic regression to model the distribution of $Z$:
\begin{equation} \label{mult}
\log\left(\frac{P(Z_i = z | {\bf x})}{P(Z_i= \text{``00''} | {\bf x})} \right)={\bf x}_i^T {\bm \gamma_{z:\tau}}, \quad i = \{1, \ldots, n\}, \quad z = \{\text{``11'',``01'',``10''}\}.
\end{equation}

The predicted probabilities, $\{p_{00}({\bf x}), p_{11}({\bf x}), p_{01}({\bf x}), p_{10}({\bf x})\}$, are then combined (equation \ref{phi}) to compute an estimate $\hat{\phi}({\bf x})$ of the conditional correlation coefficient. By plotting ${\bf x}$ versus $\hat{\phi}({\bf x})$, it is possible to show how the correlation structure depends on covariates. The estimates can be compared with the three limit values shown in Section \ref{sec:2}. Note, however, that such limits may be surpassed at some values of ${\bf x}$, due to a combination of finite-sample variability and model misspecification.

The proposed estimator is implemented in two steps, where the first step requires estimating a quantile regression model on each response separately, and the second step consists of a multinomial regression model applied to the joint distribution of the signs of the residuals. Both estimators are supported by most standard software. The variability associated with the first-step estimation of quantile regression coefficients, $\hat{\bm \beta}^{(1)}_{\tau}$ and  $\hat{\bm \beta}^{(2)}_{\tau}$, must be taken into account to evaluate correctly the  variance of the second-step estimator $\hat{\bm\gamma}_{z:\tau}$. In principle, one could use well-known results on two-step estimators (\citealp{Murphy, Hardin}). However, bootstrap is commonly used for inference on quantile regression, and represents a very convenient approach in the current framework.

\section{Dependence between lung function measures} \label{sec_ex}
\vspace{0.5cm}

Spirometric indexes are used to assess lung function impairment \citep{LungfunSite}, and the diagnosis of many pulmonary diseases is based on comparing observed measures with the tails of the distribution in the healthy population \citep{StanoLung}. Analyzing and interpreting percentiles of spirometric indexes allows identifying risk factors of respiratory impairment, diagnosing pulmonary diseases, and selecting appropriate treatments \citep{BottaiPistelliPercentiles}.

We investigated the effect of a variety of predictors on two important spirometric indexes: forced vital capacity (FVC, expressed in liters), and forced expiratory volume in one second (FEV1, expressed in liters). FVC is a measure of the volume change in the lung between a full inspiration to total lung capacity, and a maximal expiration to residual volume. FEV1 represents the volume exhaled during the first second of a forced expiratory maneuver started from the level of total lung capacity. We used data from a sample of 945 subjects from the Po river delta study \citep{Carrozzi90}, a prospective study conducted to investigate obstructive pulmonary diseases in the general population of a rural area in northern Italy. The patients' age ranged between 18 and 64 years. We only analyzed males, which represented about forty-nine percent (466 subjects) of the entire dataset. We considered four covariates: height (cm), age (years), an indicator of comorbidities such as asthma, cough or wheeze, and an indicator of smoking (0 = never smoker, 1 = ever smoker). All considered predictors are known relevant determinants of lung function (\citealp{LungfunSite}, \citealp{StanoLung}).

\begin{figure}[h]
\begin{center}
\includegraphics[width=8cm, height=8cm]{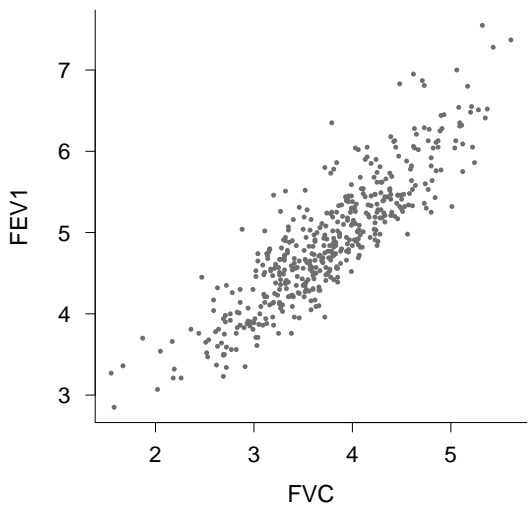}
\caption{Scatterplot of the sample values of FEV1 (liters) against FVC (liters).}\label{association_fig1}
\end{center}
\end{figure}

Exploratory analyses were performed to assess the relationship between the responses considered. Figures~\ref{association_fig1} suggested a strong positive association between FVC and FEV1 (Spearman's correlation = 0.878). 

We estimated the following quantile regression models for $\tau = \{0.1, 0.5, 0.9\}$:
\begin{equation}\label{qr1}
Q_{\text{FVC}}(\tau) = \beta^{(1)}_{\tau, 0} + \beta^{(1)}_{\tau, 1} (\text{age} - 37) + \beta^{(1)}_{\tau, 2} (\text{height} - 172) + \beta^{(1)}_{\tau, 3} \text{comorb.} + \beta^{(1)}_{\tau, 4} \text{smoke}, 
\end{equation}
\begin{equation} \label{qr2}
Q_{\text{FEV1}}(\tau) = \beta^{(2)}_{\tau, 0} + \beta^{(2)}_{\tau, 1}  (\text{age} - 37) + \beta^{(2)}_{\tau, 2} (\text{height} - 172) + \beta^{(2)}_{\tau, 3} \text{comorb.} + \beta^{(2)}_{\tau, 4} \text{smoke}. 
\end{equation}
 The estimated quantile regression coefficients are reported in Table ~\ref{table:coef1}. Standard errors were obtained from 1000 tilted bootstrap replicates (\citealp{DiCiccioRomano}; \citealp{EfronTibshirani}). Results suggested that age and height were two important predictors of lung function.

\begin{table}[h]
\centering
\small
\begin{tabular}{l|cc|cc|cc}
\hline
&  \multicolumn{2}{c|}{$\tau=0.10$} & \multicolumn{2}{c|}{$\tau=0.50$} &  \multicolumn{2}{c}{$\tau=0.90$}\\
& Coefficient  & SE & Coefficient  & SE & Coefficient  & SE  \\
\hline
 FVC&  &  &  &&  &   \\
  \hline
Intercept    &4.227    &   0.153$^*$   &4.977   & 0.118$^*$  &5.816   & 0.14$^*$    \\
Age$-37$       &-0.019  & 0.004$^*$     &-0.024    & 0.003$^*$   &-0.022   & 0.004   \\
Height$ - 172$  &0.058   & 0.008$^*$  &0.056   & 0.005$^*$  & 0.065  & 0.008$^*$  \\
Comorbidity  &0.055   & 0.111  &0.112    & 0.088 & 0.162    & 0.129   \\
Ever Smoker  &0.037  &  0.164 &-0.105  & 0.129 & -0.259    & 0.149   \\
   \hline
\hline

 FEV1&  & &  &  &   &   \\
  \hline
Intercept      & 3.168 & 0.123$^*$ & 3.961    & 0.080$^*$ & 4.413   & 0.073$^*$  \\
Age$-37$       & -0.027  & 0.003$^*$  & -0.027  & 0.003$^*$ & -0.029  & 0.003$^*$  \\
Height$ - 172$  & 0.045   & 0.006$^*$ & 0.043  & 0.004$^*$ & 0.043  & 0.006$^*$   \\
Comorbidity    & -0.188  & 0.103  & -0.029  & 0.073 & 0.044   & 0.106  \\
Ever Smoker         & 0.059    & 0.128  & -0.183 & 0.088$^*$ & -0.113  & 0.085  \\
   \hline
\end{tabular}
\caption{Estimated quantile regression coefficients with response FVC (top table) and FEV1 (bottom table). The asterisk ($^*$) indicates p-values less than $0.05$.}\label{table:coef1}
\end{table}

Using the estimated quantile regression residuals, we calculated the concordance indicator $z_i$ defined in (\ref{zed}), and modeled its conditional distribution using a multinomial regression:
\begin{equation}\label{ML}
\log\left(\frac{P(Z=z)}{P(Z=\text{``00''})} \right)= \gamma_{z:\tau, 0} + \gamma_{z:\tau, 1} s(\text{age}) + \gamma_{z:\tau, 2} s(\text{height}) + \gamma_{z:\tau,3} \text{comorb.} + \gamma_{z:\tau,4} \text{smoke}, 
\end{equation}
$z = \{$``11'', ``01'', ``10''$\}$. In the regression equation, $s(x)$ denotes the basis of a natural cubic splines with two internal knots at the empirical tertiles. Using splines allows achieving any desired flexibility, but makes it difficult to interpret the $\gamma$ model coefficients. Note, however, that the second-step multinomial regression is only used for prediction purposes. We remark that, in general, it is possible to use different sets of covariates in the first- and second-step model.

\begin{figure}[h]
\begin{center}
\includegraphics[width=12cm, height=7cm]{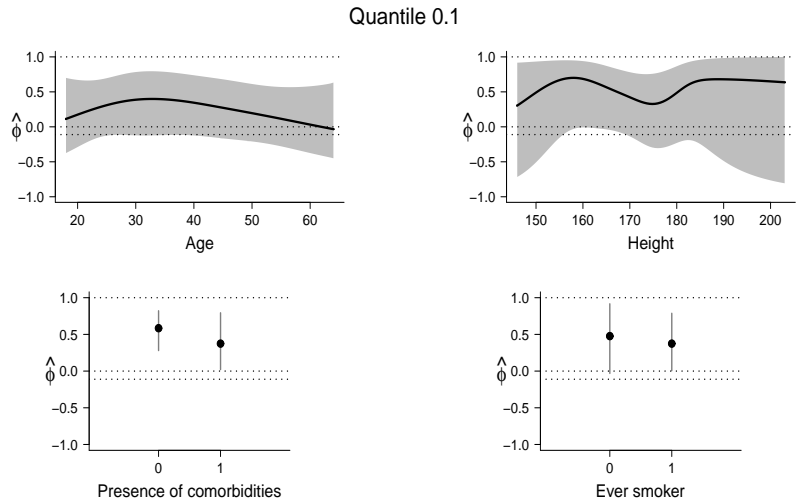}
\caption{Predicted correlation at the 10th percentile, expressed as a function of the predictors. The horizontal lines indicate, from top to bottom, perfect positive correlation ($\phi_{\text{Max}}$), independence ($\phi_{\text{Indep}}$), and perfect negative correlation ($\phi_{\text{Min}}$), respectively.}\label{prediction_fig1}
\end{center}
\end{figure}

\begin{figure}[h]
\begin{center}
\includegraphics[width=12cm, height=7cm]{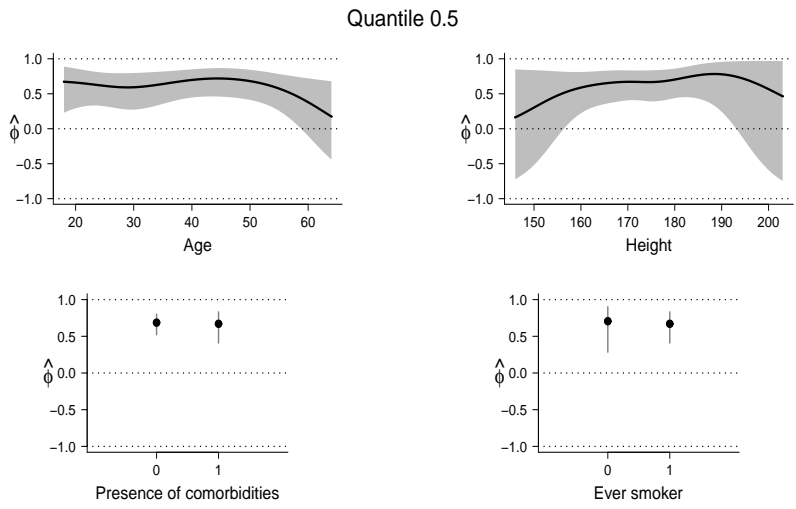}
\caption{Predicted correlation at the median.}\label{prediction_fig2}
\end{center}
\end{figure}

\begin{figure}[h]
\begin{center}
\includegraphics[width=12cm, height=7cm]{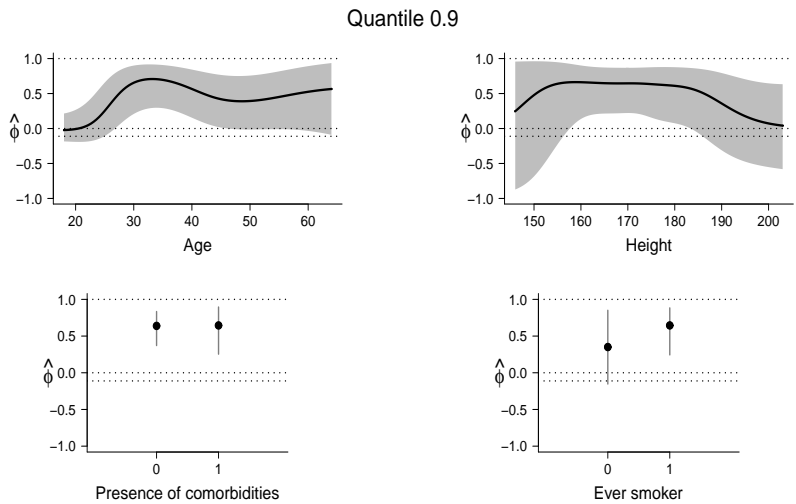}
\caption{Predicted correlation at the 90th percentile.}\label{prediction_fig3}
\end{center}
\end{figure}

The conditional correlation $\hat\phi({\bf x})$ was computed by applying equation \eqref{phi} to the fitted probabilities from model \eqref{ML}.
In Figures ~\ref{prediction_fig1}, ~\ref{prediction_fig2} and ~\ref{prediction_fig3}, we represent how correlation depends on covariates at different values of $\tau$.
To compute confidence intervals, we obtained bootstrap standard errors of $\text{logit}\left(\hat\phi({\bf x})\right)$. The values corresponding to $\phi_{\text{Indep}}$, $\phi_{\text{Min}}$, and $\phi_{\text{Max}}$ are shown as horizontal lines. 

Results showed a consistently positive correlation between FVC and FEV1. At $\tau = 0.1$, the estimated $\phi$ coefficients was considerable (close to 0.5), suggesting that patients are more often below or above the lower limit of normality with respect to both spirometric measurements. Interestingly, the correlation was slightly smaller in presence of comorbidities. This could be explained by the fact that some comorbidities may only affect one of the two response variables of interest, breaking the existing correlation. At $\tau = 0.5$, correlations were generally large, and unaffected by predictors. At $\tau = 0.9$, the estimated $\phi$ coefficient was typically around 0.5, but approached zero at young ages ($< 25$ years), and in tall patients ($> 180$ cm). This could be explained by the fact that particularly large values of both FVC and FEV1 indicate good health, without underlying pathological conditions that may induce correlation.

\clearpage
\section{National Merit Twin Study}\label{NMSQT}
In this section we consider an application to the National Merit Twin Study, that was previously analyzed, among others, by \cite{Loehlin1976} and \cite{LiChengFine}. Extensive questionnaires were administered to 839 adolescent twins identified among the roughly 600,000 US high school juniors who took the national merit scholarship qualifying test (NMSQT) in 1962. The dataset is available in the \texttt{mdhglm} R package \cite{mdhglm}, and includes 768 pairs of same-gender twins. The twins were classified as identical or fraternal based on a mail-in questionnaire. 
The NMSQT consists of five subtests, covering the domains of English, mathematics, social science, natural science, and vocabulary. A total score is calculated as the sum of the scores obtained from the five subtests. In our analysis, the bivariate outcome ($Y^{(1)}$, $Y^{(2)}$) is given by the total NMSQT scores of the twin pair. We considered the following binary covariates: Sex (0 = male, 1 = female), Income (an indicator of family income level being above 10,000 US dollars), Education (an indicator of whether at least one of the parents had education beyond high school), and Zygosity (1 for identical twins, and zero otherwise). The aim of the analysis was to study the association between twins in terms of academic abilities, conditional on the observed factors. 

We first estimated the univariate quantile regression to model covariate effects on the marginal distribution of the NMSQT scores ($Y^{(1)}$, $Y^{(2)}$), through a linear quantile regression model ($\tau = 0.01, 0.02, \ldots 0.99$) that also included an interaction term Zigosity$\times$Income. Results showed that male students
coming from families with high income and high parental education tend to score higher. The interaction was generally significant, suggesting that, in whealthier families, identical twins tend to perform better than heterozygous twins.

We then calculated the concordance indicator $z_i$ defined in (\ref{zed}), and estimated the multinomial model of concordance.
Following \cite{LiChengFine}, we restricted the analysis to quantiles in the range $[0.2, 0.8]$. Because NMSQT scores in each twins' pair can be considered exchangeable (as the twins cannot be ordered), the discordant residuals can be considered as a single category, simplifying the multinomial regression model. We initially included the same predictors used in the univariate quantile regression models; however, our final second-step model only included zigosity, that was the only significant predictor.

\begin{figure}[h]
\begin{center}
\includegraphics[width=12cm, height=7cm]{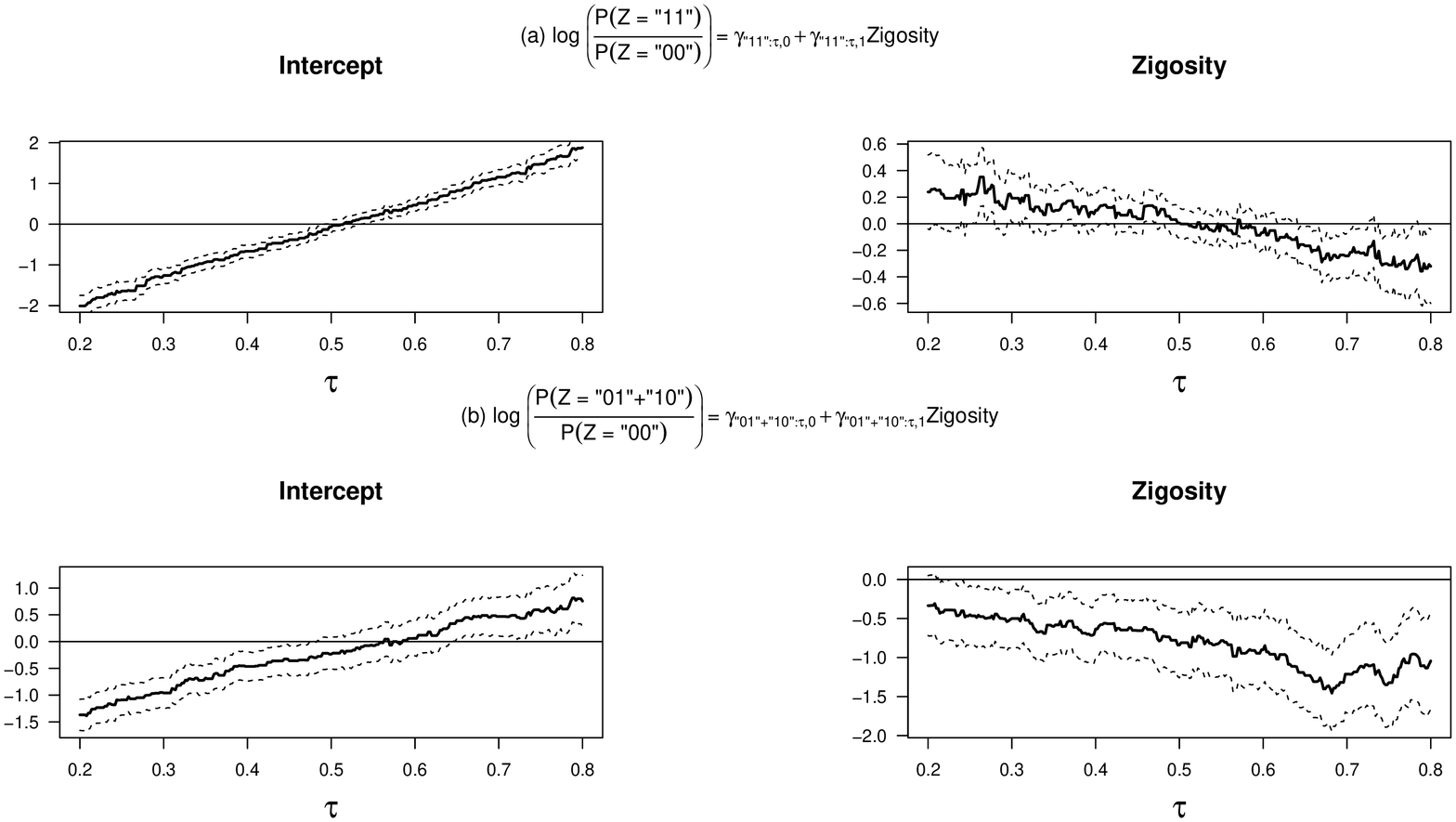}
\caption{Estimated coefficients $(\hat{\gamma}_{z:\tau})$ of the multinomial logistic model, $\tau \in [0.2,0.8]$.}\label{Mult}
\end{center}
\end{figure}

Figures ~\ref{Mult} (a) and (b) illustrate the parameters' estimates ($\hat{\gamma}_{z:\tau}$) together with 95\% bootstrap confidence intervals. The figures compare the negative concordance (Z =``11'') and the discordance (Z =``01'' + ``10'') to the situation of positive concordance (Z =``00''). The model coefficients show that identical twins have a lower chance of being discordant and a higher chance of being positively concordant. The estimated coefficients are decreasing function of the quantile, showing a stronger difference between twins with higher performances in NMSQT test. 

\begin{figure}[h]
\begin{center}
\includegraphics[scale=0.5]{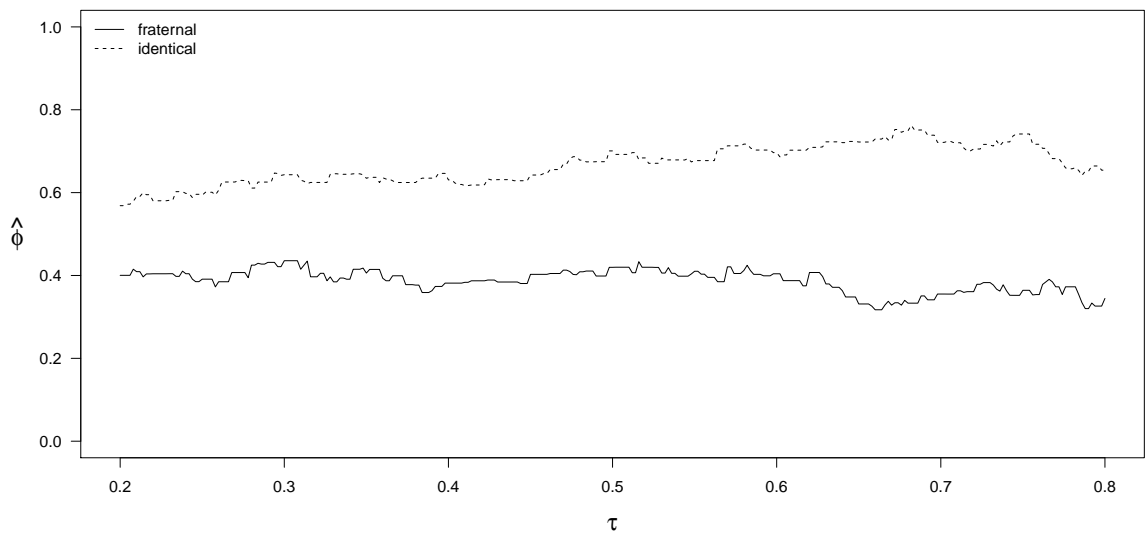}
\caption{Predicted correlation $\hat\phi({\bf x})$ for fraternal twins and identical twins, $\tau \in [0.2,0.8]$.}\label{zigosity}
\end{center}
\end{figure}

In Figure \ref{zigosity} we report the estimates of the conditional coefficient of correlation $\hat\phi({\bf x})$. In the calculation of $\hat\phi({\bf x})$, the predicted probability of discordant residuals in equation \eqref{phi} was equally split in the two terms $p_{01}$ and $p_{10}$. The correlation was always positive for both identical and fraternal twins, and was higher for identical ones. These results are in line with those obtained by \cite{LiChengFine}, and strongly support the presence of a genetic component in the students' performances.

\clearpage
\section{Final remarks} \label{sec:conclusions}
\vspace{0.5cm}

Dependence between multiple response variables of interest is very common, and occurs in cross-sectional, case-control, and longitudinal studies. The method presented in this paper allows to investigate the conditional correlation structure of a multivariate response, by analyzing the signs of the residuals from univariate quantile regression models. In the paper, we assumed that the same quantile, $\tau$, was estimated for all outcomes. However, depending on the purpose of the analysis, one could consider a different quantile, say $\tau^{(j)}$, for each outcome of interest, $j = 1, \ldots, d$.

In principle, other families of regression models could be used in place of quantile regression: for example, one could apply the same method to the residuals of a linear (mean) regression. However, the residuals' signs are a natural outcome of quantile regression, in which by definition a proportion $\tau$ of the observations lie below the estimated regression line. Also, in equation (\ref{q}), the conditional quantiles of $Y^{(j)}$ are assumed to be linear in the parameters. Although this parametrization is very popular and computationally convenient, the method presented in this paper can be applied to any nonlinear quantile function. 

In our work, we mainly considered a bivariate response. When the response vector has more than two elements, the proposed approach can be modified by introducing higher-order sign-concordance probabilities, that can be combined into more complicated summary statistics. For example, the simple correlation coefficient $\phi$ defined in equation (\ref{phi}) could be replaced by some multivariate measure of correlation (e.g. \cite{Wang2020}).

The used multinomial logistic model is completely unstructured, and has a number of parameters which is proportional to $2^d$ where $d$ is the number of outcomes being considered. Some simplifying assumptions may be used to investigate large-dimensional binary responses. In the case $d = 2$, one could directly model a binary variable indicating concordance ($Z_i = \{00, 11\}$) and discordance ($Z_i = \{01, 10\}$), using standard logistic regression.

All methods described in this paper are implemented in standard software. The R code used to analyze the two datasets presented in Sections \ref{sec_ex} and \ref{NMSQT} is available upon request to the authors.

\section*{Acknowledgments}
We thank Dr. Giovanni Viegi for allowing use of a subset of the data from the Po river delta epidemiological study.


\end{document}